\documentclass{revtex4}
\usepackage{epsf}

\begin{document}
\title{Dineutron structure in $^{8}$He}

\author{Yoshiko Kanada-En'yo}
\address{Yukawa Institute for Theoretical Physics, Kyoto University,
Kyoto, Japan}

\begin{abstract}
The ground and excited states of $^{8}$He were investigated with a method of 
antisymmetrized molecular dynamics(AMD). 
We adopted effective nuclear interactions which systematically reproduce 
the binding energies of $^4$He, $^6$He and $^8$He. The ground state of 
$^8$He has both the $j$-$j$ coupling feature($p_{3/2}$ closure) and 
the $L$-$S$ coupling feature($^4$He$+2n+2n$) 
with a slight tail of dineutron at the long distance region.
The theoretical results give an indication of the $0^+_2$ state with 
dineutron gas-like structure. The dineutron structure,
$^4$He+$2n$+$2n$, of this state is
similar to the $3\alpha$-cluster structure of the $^{12}$C($0^+_2$) state 
which has been interpreted as an
$\alpha$ condensate state.
Since the $^8$He($0^+_2$) state has a significant overlap with the 
dineutron condensate wave function where
two dineutrons are moving in
$S$ wave around the $\alpha$ core with a dilute density, 
we suggest that this theoretically predicted $0^+_2$ state is a candidate
of the dineutron condensate state.

\end{abstract}
\maketitle

\noindent

\section{Introduction}	

In the recent progress of unstable nuclear physics, 
various kinds of exotic structure have been discovered.
Many of these phenomena in light nuclear region
are often related to cluster physics. 
From the viewpoints of nuclear cluster,
there are many theoretical works on halo structure in neutron-rich nuclei
and molecular structure in Be isotopes. 
Recently, Tohsaki {\em et al.} proposed a new type of cluster structure
in the second $0^+$ state $^{12}$C, where 
3 $\alpha$ clusters are weakly interacting\cite{Tohsaki01}. 
This is a dilute gas state of 
$\alpha$ particles which behave as bosonic particles in the dilute density.
This phenomena is associated with 
Bose-Einstein Condensation(BEC) and is called ``alpha condensation''.
The alpha condensation was originally 
suggested in dilute nuclear matter by R\"opke et al.\cite{Ropke98}.
The $0^+_2$ of $^{12}$C is regarded as an example, 
where the alpha condensation 
is realized in a finite nuclear system.
Then, it is challenging to search for 
such cluster-gas states in other nuclei.
In analogy to the alpha condensation, 
dineutron condensation in neutron matter is a recent key issue in physics of 
unstable nuclei. Matsuo suggested that the dineutron 
correlation can be enhanced in dilute neutron matter\cite{Matsuo06}. 
In real systems, one should focus on dineutron correlation in finite nuclei
such as halo nuclei and extremely neutron-rich nuclei, or that in neutron
skin at a surface region of neutron-rich nuclei.
In fact, the dineutron correlation in two-neutron halo nuclei like 
$^6$He and $^{11}$Li
attracts great interests in these days. In case of $^6$He, where the 
$^4$He is the good core, the dineutron correlation of the valence neutrons
has been demonstrated in three-body model calculations
(for example, \cite{Bertsch91,Zhukov93,Aoyama01,Arai01} 
and references therein).

Now, let us consider structure of $^8$He from a point of view
associated with the dineutron condensation.
Firstly, more than one dineutrons are required to 
construct a dineutron condensate state. 
In $^8$He, two pairs of neutrons are possible from four valence neutrons 
around the $^4$He core.  
In second, $^8$He system may have some correspondence with the 
$^{12}$C system, because both of them have the same neutron number, $N=6$.
In analogy to $^{12}$C, the ground state of $^8$He 
may have a feature of the neutron $p_{3/2}$ closure 
or the SU(3)-limit $p$-shell configuration.
Instead of the ground state, one can speculate the dineutron gas-like state
with developed $^4$He+$2n$+$2n$ structure in excited states.

There are many theoretical works on He isotopes. 
Application of {\it ab initio} calculations such as GFMC and NCSM
with realistic nuclear forces have now reached to the mass $A\sim 10$ region
including $^6$He and $^8$He\cite{Pieper04,Pieper05,Caurier06}.
Systematic studies of He isotopes have been performed 
also by model calculations with effective interactions
such as cluster models as well as 
GSM\cite{Michel03,Volya05,Hagen05} 
and mean field approaches\cite{Sugahara96}.
Three-body model with an assumption of the $^4$He core
has been often adopted to study $^6$He 
\cite{Bertsch91,Zhukov93,Aoyama01,Arai01,Csoto93,Baye94} and it has been 
applied to heavier He isotopes \cite{Aoyama02}.
$^8$He and $^{10}$He have been also studied by such 
models as $^4$He+$Xn$ models \cite{Suzuki90,Varga94,Itagaki00,Masui07} 
and by extended models \cite{Dote00,Aoyama06} 
which have less assumption of the core.
With Fermionic molecular dynamics, the study of He isotopes  
has been performed based on a realistic nuclear force \cite{Neff05}.
However, many of these studies are concentrated on the ground states
except for three-body models, GSM  and GFMC.

After the experimental indication of neutron skin structure in 
$^8$He\cite{Tanihata92}, many experimental works 
on $^8$He have been recently 
performed to reveal the detailed properties of the 
ground state.
The core excitation 
$^6$He$(2^+)$ in the ground state, which has been experimentally 
suggested\cite{Korsheninnikov03},
indicates that $^8$He is different from a simple three-body
state of $^6$He$(0^+)$+$2n$.
Recent experiments using $^8$He beams suggested 
the significant component of the $(p_{3/2})^2(p_{1/2})^2$ 
configuration \cite{Chulkov05,Keeley07}. They may
support dineutron correlation in the $^8$He ground 
state rather than the pure $(p_{3/2})$ closure of neutrons.
On the other hand, a measurement of spectroscopic factor 
of $^7$He($3/2^-$)\cite{Skaza06} in $^8$He 
suggested the pure sub-shell closed structure 
contradictory to the other experimental results.
Thus, the neutron structure of the $^8$He ground state is 
controversial. Concerning excited states, 
although some levels are known to exist in the energy $E_x=3\sim 8$ 
MeV region, the experimental information is very poor for 
these states except for the $2^+_1$ state \cite{Korsheninnikov93}.

In this paper, we investigated structure of $^8$He. 
In particular, we focused on $0^+$ states and discuss their 
dineutron component, because
one of our major aims is to search for the dineutron gas-like state.
We applied a method of 
antisymmetrized molecular dynamics(AMD)\cite{ENYObc,ENYOsup,AMDrev}, 
which 
has been already proved to be 
useful in describing cluster structure in light nuclei.
AMD has been applied to 
various light unstable nuclei such as He, Li, Be isotopes as well as 
stable nuclei. It has been applied also for study of 
cluster gas-like states in $^{12}$C and $^{11}$C($^{11}$B)\cite{Enyo-c12v2,Enyo-c11}.
In the present work, we adopted 
a AMD+generator coordinate method(GCM). 
Namely, we superposed a number of AMD wave functions, which were obtained 
by energy variation with constraints,  
to take various configurations into account.
We comment that the theoretical method AMD+GCM of the present calculation 
is similar to those of the 
AMD+GCM and AMD+SSS works on He isotopes 
by Itagaki and his collaborators \cite{Itagaki00,Aoyama06} in a sense 
that multi configurations of AMD wave functions are superposed.
In \cite{Itagaki00,Aoyama06}, $^4$He+$Xn$ and 
$t+t+Xn$ configurations were {\it a priori} assumed. Another claim is that 
they used an effective interaction which makes a bound $^2n$.
In the present work, we have no assumption of the cluster core 
and chose effective interactions 
by taking care of subsystem energies such as $\alpha$-$n$ and $^6$He 
as well as nucleon-nucleon scattering.
We used some sets of interaction parameters and showed 
the calculated results of the ground and excited states of He isotopes.
By assuming $(0s)^2$ configuration as the interior structure of a 
dineutron, we analyzed dineutron structure of $^8$He and compared it
with the $\alpha$-cluster structure of $^{12}$C.

The paper is organized as follows. In the next section, we briefly explain 
the theoretical method of the present work. Results are given in 
\ref{sec:results}, and dineutron structure is discussed in
\ref{sec:discussions}.
Finally, we give a summary in \ref{sec:summary}.

\section{Formulation} \label{sec:formulation}
In this section, we briefly explain the formulation of AMD+GCM 
in the present calculation.
The detailed formulation of the AMD method 
for nuclear structure study is described in 
\cite{ENYOsup,AMDrev}.
There are various versions of practical methods of the AMD framework.
In the present work, we performed superposition of 
a number of AMD wave functions obtained 
by energy variation with constraints based on the concept of GCM.
The procedure of the variation, spin and parity projection  
and superposition is similar to those of AMD+GCM calculations in
\cite{Itagaki00,Kimura04,Enyo04},
though the details of model wave functions and effective interactions
are different from each other.

An AMD wave function is a Slater determinant of Gaussian wave packets;
\begin{equation}
 \Phi_{\rm AMD}({\bf Z}) = \frac{1}{\sqrt{A!}} {\cal{A}} \{
  \varphi_1,\varphi_2,...,\varphi_A \},
\end{equation}
where the $i$th single-particle wave function is written by a product of
spatial($\phi$), intrinsic spin($\chi$) and isospin($\tau$) 
wave functions as,
\begin{eqnarray}
 \varphi_i&=& \phi_{{\bf X}_i}\chi_i\tau_i,\\
 \phi_{{\bf X}_i}({\bf r}_j) &=& \left(\frac{2\nu}{\pi}\right )^{\frac{3}{4}} 
\exp\bigl\{-\nu({\bf r}_j-\frac{{\bf X}_i}{\sqrt{\nu}})^2\bigr\},
\label{eq:spatial}\\
 \chi_i &=& (\frac{1}{2}+\xi_i)\chi_{\uparrow}
 + (\frac{1}{2}-\xi_i)\chi_{\downarrow}.
\end{eqnarray}
$\phi_{{\bf X}_i}$ and $\chi_i$ are spatial and spin functions, and 
$\tau_i$ is isospin
function which is fixed to be up(proton) or down(neutron). 
The width parameter $\nu$ is chosen to be the optimum value for 
each system. Accordingly, an AMD wave function
is expressed by a set of variational parameters, ${\bf Z}\equiv 
\{{\bf X}_1,{\bf X}_2,\cdots, {\bf X}_A,\xi_1,\xi_2,\cdots,\xi_A \}$.

The energy variation was performed for the parity-projected 
AMD wave function $\Phi^\pm_{\rm AMD}({\bf Z})$ under constraints.
In order to obtain basis wave functions,
we adopted the total oscillator quanta and deformation
as the constraints. Hereafter, we note the 
expectation value of an operator $\hat O$ with respect 
to a normalized parity-projected AMD wave function as
$\langle \hat O \rangle$.
Expectation values 
$\langle \hat N^{\rm ho} \rangle$ of the total oscillator quanta is 
given by the
creation and annihilation operators of harmonic oscillator 
in the same way as
\cite{Enyo04}. 
In the AMD+GCM calculations with the $\beta$-constraint
(for example \cite{Kimura04}), 
the deformation is usually constrained by using the rotational invariant 
value $D\equiv Tr(QQ)/Tr^2(Q)$, 
where the matrix $Q$ is calculated by 
quadrupole operators as $Q_{\sigma\rho}=\langle \sum_i \hat\sigma_i
\hat\rho_i \rangle$ ($\hat\sigma=\hat{x},\hat{y},\hat{z}$ and 
$\hat\rho=\hat{x},\hat{y},\hat{z}$) \cite{Dote97}. 
Here $D$ is approximately related to the quadrupole 
deformation parameter $\beta$ as $D(\beta)= (5\beta^2/2\pi+1)/3$. 
In the present work, we used the modified quadrupole matrix 
$Q'_{\sigma\rho}\equiv Q_{\sigma\rho}-A\delta_{\sigma\rho}$
($A$ is the mass number)  
instead of the original $Q_{\sigma\rho}$ and imposed the constraint on the 
$D'\equiv Tr(Q'Q')/Tr^2(Q')$. This is useful for He isotopes to obtain 
basis wave functions with various configurations 
on mesh points of the two-dimensional 
parameters, $\beta$ and $\langle \hat N^{\rm ho} \rangle$.
The energy variation with 
the constraint values $N_{\rm const}$ and $\beta_{\rm const}$
was performed with respect to 
the parity-projected AMD wave function
by minimizing the energy defined as, 
\begin{equation}
E\equiv \langle \hat{H}\rangle+
V^{N}(N_{\rm const}-\langle \hat N^{\rm ho} \rangle)^2
+V^{\beta}(D(\beta_{\rm const})-D')^2.\label{eq:const}
\end{equation}
Here the artificial potentials are introduced to satisfy the condition of the 
constraints.
With a given set of constraint values $(N_{\rm const},\beta_{\rm const})$
the optimum wave function $\Phi^\pm_{\rm AMD}(N_{\rm const},\beta_{\rm const})$
was obtained. Finally, we superposed the spin-parity eigen states projected 
from the obtained wave functions, 
\begin{equation}
|^8{\rm He}(J^\pm_n)\rangle = 
\sum_{N_{\rm const},\beta_{\rm const}}c^{J\pm}_n(N_{\rm const},\beta_{\rm const})
|P^{J}_{MK}\Phi^\pm_{\rm AMD}(N_{\rm const},\beta_{\rm const})\rangle,
\end{equation}
where the coefficients $c^{J\pm}_(N_{\rm const},\beta_{\rm const})$
were determined 
by diagonalizing the Hamiltonian and Norm matrices. In the present 
calculations, we took only $M=K=0$ states.

\section{Results}\label{sec:results}
\subsection{Calculations}
$^6$He, $^8$He and $^{10}$He were calculated by the AMD+GCM method.
 The strengths, 
$V^{N}$ and $V^{\beta}$, for the constraint potentials in eq.\ref{eq:const}  
are chosen to be 30 MeV and 
2000 MeV, respectively. 
We chose the width parameter $\nu$ to optimize the energy 
for the $P^{J=0}_{(MK)=(00)}\Phi^+_{\rm AMD}(N_{\rm const}
=N_{\rm min}+2)$, which gives the minimum energy among the states
$P^{J=0}_{(MK)=(00)}\Phi^+_{\rm AMD}(N_{\rm const})$ 
in most cases. Here, $N_{\rm min}$ is 
the minimum value of the harmonic-oscillator quanta, 
$N_{\rm min}=$2, 4, and 6 for $^6$He, $^8$He, and $^{10}$He, respectively.
A common $\nu$ value for each He isotope are 
used in the calculation with each interaction.
The adopted $\nu$ values 
are listed in table \ref{tab:int}.
We adopted the constraint values of the mesh points $(i,j)$ on the
$N_{\rm const}$-$\beta_{\rm const}$ plane as
$N^{(i)}_{\rm const}=N_{\rm min}+\Delta^{(i)}$($\Delta^{(i)}$=0,1,2,3,4,6,8,10 
for positive parity states and 
$\Delta^{(i)}$=1,2,3,4,6,8,10 for negative parity states)
and $\beta^{(j)}_{\rm const}$=0, 0.2, 0.4, 0.6, $\cdots$, 1.6.  
Then, the total number of the basis wave functions are 72(63) for 
positive(negative)-parity 
states. On the $N_{\rm const}$-$\beta_{\rm cont}$ plane,
we first obtained the wave function
$\Phi^\pm_{\rm AMD}(N_{\rm const},\beta_{\rm const})$ at
$N_{\rm const}=N_{\rm min}+2$ and 
$\beta_{\rm const}$=0, 0.2, 0.4, 0.6, $\cdots$, 1.6. Then we searched for 
$\Phi^\pm_{\rm AMD}(N_{\rm const}+1,\beta_{\rm const})$ 
(or $\Phi^\pm_{\rm AMD}(N_{\rm const}-1,\beta_{\rm const})$) 
starting from 
the $\Phi^\pm_{\rm AMD}(N_{\rm const},\beta_{\rm const})$
by increasing(or decreasing) $N_{\rm const}$ one by one. 

Some of the basis wave functions with the constraints have the breaking of 
the $^4$He-core. Such the basis wave functions 
with the $^4$He-core breaking have high energies in general, 
and therefore, they
practically give only small contribution to the 
low-lying states of $^6$He, $^8$He and $^{10}$He isotopes.
It means that the $^4$He cluster is a rather good core in 
$^6$He, $^8$He and $^{10}$He isotopes, while the motion of valence neutrons
is relatively important. 
 
\begin{table}[ht]
\caption{ 
\protect\label{tab:int} Parameter sets of the effective interaction
and the values of width parameter $\nu$ adopted in the present work. 
The theoretical values of scattering length $a_s$($a_t$) 
for singlet(triplet) even channel,  
neutron separation energy of 
$^5$He ($S_{n}(^5$He)$\equiv E(^4$He)$-E(^4$He-$n$)), 
$2\alpha$ threshold energy of $^8$Be, 
two-neutron separation energies of $^6$He and $^8$He 
($S_{2n}(^6$He)$\equiv E(^4$He)$-E(^6$He) and 
$S_{2n}(^8$He)$\equiv E(^6$He)$-E(^8$He))
are also listed. 
} 
\begin{tabular}{@{\qquad}c@{\qquad}c@{\qquad}c@{\qquad}c@{\qquad}c@{\qquad}c}
\hline\hline
Parameter set & &  v58 & v56 & m62 & m56 \\
\hline 
Central force & & Volkov No.2 & Volkov No.2 & MV1 case(3) & MV1 case(3)\\
\hline
Wigner & $w$ & 0.42 & 0.44 & 0.38 & 0.44 \\
Bartlett & $b$ & 0 & 0.15 & 0 & 0.15 \\
Heisenberg & $h$ & 0 & 0.15 & 0 & 0.15 \\
Majorana & $m$ & 0.58 & 0.56 & 0.62 & 0.56 \\
\hline
$\nu(^4$He) (fm$^{-2}$) & &0.265 & 0.265 & 0.210 & 0.210 \\
$\nu(^6$He) (fm$^{-2}$) & &0.245 & 0.245 & 0.210 & 0.210 \\
$\nu(^8$He) (fm$^{-2}$) & &0.240 & 0.240  & 0.185 & 0.185 \\
$\nu(^{10}$He) (fm$^{-2}$) & &0.185 & 0.175 & 0.165& 0.165\\
\hline
\hline
 & exp. &  v58 & v56 & m62 & m56 \\
\hline
$a_t$ (fm) & 5.42 ($p$-$n$) & 9.7  & 5.4  & 6.4  & 4.2 \\
$a_s$ (fm) & $-$16.5 ($n$-$n$) & 9.7  & $-$23.9  & 6.4  & $>$100 \\
$S_{n}(^5$He) (MeV) & $-$0.9  & $-$0.7  & $-$0.7  & $-$1.0  & $-$0.4\\ 
$2E(^4$He)$-E(^4$He-$^4$He) (MeV) & $-$0.1  & 0.6  & 1.4  & $-$1.3  & $-$0.6 \\
$S_{2n}(^6$He) (MeV) & 1.0  & 1.3  & $-$0.2  & 2.1  & 1.1\\ 
$S_{2n}(^8$He) (MeV) & 2.1  & 3.0  & 3.2  & 1.2  & 2.0 \\
\hline
\hline
\end{tabular}
\end{table}

\subsection{Interactions}
We used effective nuclear interaction consisting of the 
central force, 
the spin-orbit force and Coulomb force.
As for the central force, 
we adopted the Volkov force\cite{Volkov}
used in the work on 
He isotopes with AMD+GCM($^4$He+$Xn$)\cite{Itagaki00}, 
and also the MV1 force\cite{MV1} used in the AMD calculations
of $^{12}$C \cite{Enyo-c12,Enyo-c12v2}.  
We used the spin-orbit force of the G3RS force\cite{LS} as done in 
\cite{Itagaki00,Enyo-c12}.
We fixed the strengths of the spin-orbit term as 
$u_{ls}$ = 2000 MeV, which is the same value as
in \cite{Itagaki00}.
By taking care of energies of subsystems, we tuned the interaction 
parameters, $w$, $b$, $h$, $m$,  for 
Wigner, Bartlett, Heisenberg and Majorana exchange terms
in the the central force(Volkov or MV1), respectively.
$^6$He, $^8$He and $^{10}$He were calculated with AMD+GCM by using 
totally 4 cases of central force. The parametrization for the central force
is summarized
in table \ref{tab:int}. 
In order to demonstrate characteristics of the effective interactions, 
we also show the relative energies of subsystems and the 
nucleon-nucleon scattering lengths with these 4 types of interaction.
We estimate the energies of the $^4$He, $^4$He-$n$ state with
$J^\pi=3/2^-$, and $^4$He-$^4$He state with $J^\pi=0^+$, 
by assuming the $(0s)^4$ state of $^4$He and performing 
cluster-GCM calculations within the $\alpha$-$n$ and 
$\alpha$-$\alpha$ cluster models for simplicity.

The first case of interaction 
is Volkov No.2 force\cite{Volkov} with interaction parameters 
$m=0.58$, $b=h=0$. This is the same effective interaction as that used in 
the AMD+GCM($^4$He+$Xn$) by Itagaki et al. \cite{Itagaki00}, 
which succeeded to systematically reproduce the binding energies of 
He isotopes. We note this interaction 'v58' in this paper. 
In spite of good agreement of the binding energies of He isotopes, 
the v58 force has a fault that 2 neutrons are
bound in a free space. It is well known that the Volkov force
with $b=h=0$ has 
too strong neutron-neutron attraction, because such the parametrization 
with no Bartlett term nor Heisenberg term gives 
the same interaction in the singlet-even channel as  
that in the triplet-even channel. 
In reality, the singlet-even channel 
has weaker attraction, and two neutrons are unbound. 
In order to describe dineutron correlation in neutron-rich nuclei it might be
crucial to reproduce such the feature of two-nucleon system, though 
it does not matter in case of spin-isospin saturated 
systems like $Z=N$ nuclei.

In the second case of interaction, 
we used Volkov No.2 force with modified interaction parameters 
as $m=0.56$, $b=h=0.15$. 
This interaction (noted as 'v56') describes
well the experimental $S$-wave
scattering lengths of the $n$-$n$ and $p$-$n$ channels, and the unbound
feature of 2-neutron system. The Majorana parameter 
$m=0.56$ was determined by adjusting the binding energy of $^8$He to the 
experimental data.
However, this interaction fails to reproduce 
$2n$ separation energies 
of $^6$He and $^8$He, and it also gives too strong attraction in 
$^4$He-$^4$He system. 

The third interaction('m62') and the forth one('m56') 
listed in table \ref{tab:int}
are based on the MV1 force\cite{MV1}. 
The parametrization of the m62 interaction is 
$m=0.62$ and  $b=h=0$, which is the same as used in the AMD calculations
of $^{12}$C \cite{Enyo-c12v2,Enyo-c12}. 
In case of the m62 interaction, 
two neutrons are bound in a free space
as well as the Volkov force with $b=h=0$ like the v58 interaction. 
In the 'm56' interaction, we used the modified Bartlett and Heisenberg
terms, $b=h=0.15$, and the Majorana term $m=0.56$ 
which was adjusted to reproduce 
the binding energy of $^8$He. With the $m=0.56$ interaction, 
two neutrons are almost unbound in a free space, and 
other energies of subsystems are reasonably reproduced. 

\subsection{Ground states of He isotopes}
\begin{figure}[th]
\epsfxsize=7 cm
\centerline{\epsffile{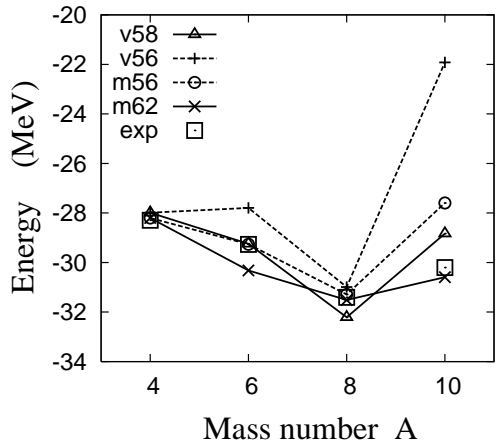}}
\vspace*{8pt}
\caption{The calculated energies of He isotopes with the v58, v56, m62 and m56
interactions(see text). The experimental data
are also given.
\label{fig:he-be}
}
\end{figure}

We show the calculated results of 
the ground states of He isotopes.
The energies of He isotopes are shown in Fig.~\ref{fig:he-be}.
The v58 and m56 interactions systematically 
reproduce the energies of $^4$He, $^6$He and $^8$He, 
though they overestimate the $^{10}$He energy.
On the other hand, the v56 and m62 interactions are poor in 
reproduction of the $^6$He energy, and therefore, they 
fail to reproduce two-neutron separation energies 
of $^6$He and $^8$He as shown in 
table \ref{tab:int}. Hereafter, we discuss the results obtained with 
the v58 and m56 interactions. We stress again that
the v58 interaction well describes the energies of subsystems except for 
the fault of the too strong neutron-neutron interaction, while
the m56 interaction reasonably reproduces the global features of 
the subsystem energies.

\begin{figure}[th]
\epsfxsize=5. cm
\centerline{\epsffile{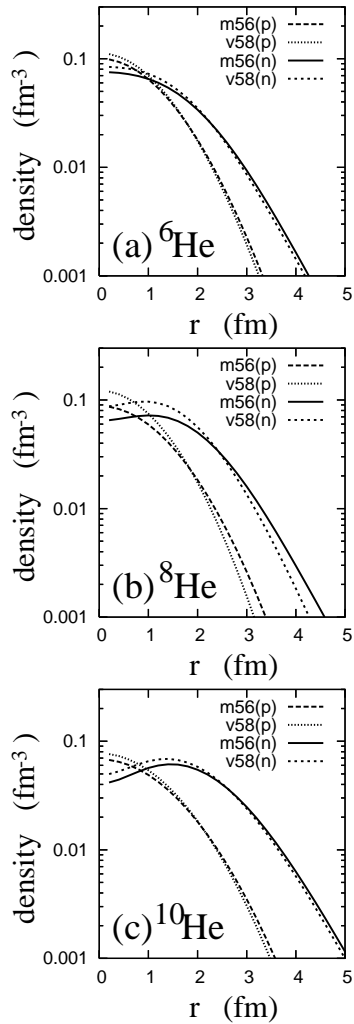}}
\vspace*{8pt}
\caption{Point-proton and point-neutron density in the ground states
of He isotopes. The calculated results are those with the m56 and v58 interactions.
\label{fig:rdens}
}
\end{figure}

The calculated 
root-mean-square radii of proton, neutron and matter density 
are given in table 
\ref{tab:rmsr} with the experimental data.
The theoretical results of 
other calculations are also listed.
Experimentally, extremely large radii of 
$^6$He and $^8$He have been reported 
by the reaction cross sections \cite{Tanihata85,Tanihata88,Tanihata92}.
It has been suggested that the large radii originate in 
the remarkable enhancement of neutron radii 
due to the neutron-halo and neutron-skin structures in $^6$He and 
$^8$He, respectively.
The empirical neutron radii are well described by the present calculations
with the m56 interaction. On the other hand, 
the neutron radii calculated with the v58 interaction 
are slightly smaller 
than the empirical ones 
as well as the former AMD+GCM($^4$He+$Xn$) calculations with the same v58 
interaction\cite{Itagaki00}.
The proton radii calculated with the m56 interaction are consistent with
the observed data except for that of $^4$He.
Figure \ref{fig:rdens} shows the proton density and neutron density.
In $^6$He, the neutron density has a long tail at a large distance region. 
This is the neutron halo structure and is similar to the neutron 
density obtained by other calculations such as SVM\cite{Varga94}.
In $^8$He, the neutron and proton density shows the neutron skin structure
at the surface, which well corresponds to the discussion 
in \cite{Tanihata92,Varga94}.
Thus, the present calculations with the m56 interaction 
systematically describe the ground-state properties of 
$^6$He and $^8$He such as energies and radii. 

Let us discuss the effect of the spin-orbit force, which may induce
the $j$-$j$ coupling feature of neutrons.
The expectation values of the spin-orbit force $\langle V_{ls} \rangle$ and 
those of the squared total intrinsic spin 
of neutrons $\langle S_n^2 \rangle$ are listed in table \ref{tab:he8ex}.
From the values of $\langle S_n^2 \rangle$,  
the $S=1$ component in the $^6$He$(0^+_1)$ state 
is estimated to be 0.13 and 0.07 in the m56 and v58 results, respectively.
It means that the $(p_{3/2})^2$ 
configuration is contained due to the spin-orbit force. 
However, the $S=0$ component is still significant 
because of $L$-$S$ coupling feature of 
spin-zero $2n$ correlation.
We note that 
the fraction 0.87 in the m56 results 
for the $S=0$ component in $^6$He 
is in good agreement with three-body model calculations
\cite{Arai01,Csoto93,Baye94,Hagino05}. 
Compared with the results of $^6$He, where the $L$-$S$ coupling configuration
is significant as well as the $j$-$j$ coupling configuration,
the $j$-$j$ coupling feature increases in 
the $^8$He$(0^+_1)$ state because of the $(p_{3/2})^4$ closure.
As a result, the spin-orbit force gives much larger
attraction in $^8$He by factor $3\sim 4$ than in $^6$He. 
It is interesting that the
the value $\langle S_n^2 \rangle=0.86$(0.72) of the $^8$He$(0^+_1)$
in the m56(v58) results is different from 
the value $\langle S_n^2 \rangle=1.33$ for the pure
$(p_{3/2})^4$ closed state. This deviation is 
because the $L$-$S$ coupling configuration
is still contained in $^8$He due to the spin-zero $2n$ correlation of neutron pairs. 
The detailed dineutron structure of $^6$He and $^8$He will be discussed 
later.

\begin{table}[ht]
\caption{ 
\protect\label{tab:rmsr} Root-mean-square radii (fm) of point-proton, point-neutron and point-matter density of the ground states of He isotopes. 
The experimental value(a) is deduced from the charge 
radius\protect\cite{Wang04}, and 
empirical values(b) are taken from \protect\cite{Tanihata92,Tanihata88}.
Theoretical values of other calculations, NCSM\protect\cite{Caurier06}, 
SVM\protect\cite{Varga94} AMD+GCM($^4$He+$Xn$)\protect\cite{Itagaki00},
RMF\protect\cite{Sugahara96} are also given. }
\begin{tabular}{ccccccccc}
\hline\hline
 &  & exp. & AMD-v58 & AMD-m56 & SVM\protect\cite{Varga94}
& RMF\protect\cite{Sugahara96} & AMD($^4$He+$Xn$)\protect\cite{Itagaki00}
& NCSM\protect\cite{Caurier06} \\
\hline
$^4$He & $r_p$ & 1.455(1) & 1.46  & 1.64  &  &  &  & 1.45\\
 & $r_n$ &  & 1.46  & 1.64  &  &  &  & 1.45\\
 & $r_m$ &  & 1.46  & 1.64  &  & 1.76 &  & \\
\hline
$^6$He & $r_p$ & 1.912(18)$^{(a)}$ & 1.83  & 1.90  & 1.80  &  &  & 1.89\\
 & $r_n$ & $2.59-2.61$$^{(b)}$& 2.40  & 2.49  & 2.67  &  &  & 2.67 \\
 & $r_m$ & $2.33-2.48$$^{(b)}$& 2.23  & 2.31  & 2.46  & 2.43  & 2.32 & \\
\hline
$^8$He & $r_p$ & $1.76-2.15$$^{(b)}$ & 1.76  & 1.96  & 1.71  &  &  & 1.88\\
 & $r_n$ & $2.64-2.69$$^{(b)}$ & 2.37  & 2.63  & 2.53  &  &  & 2.8\\
 & $r_m$ & $2.49-2.52$$^{(b)}$ & 2.24  & 2.48  & 2.40  & 2.55  & 2.31 & \\
\hline
$^{10}$He & $r_p$ &  & 2.04  & 2.13  &  &  &  & \\
 & $r_n$ &  & 2.88  & 2.97  &  &  &  & \\
 & $r_m$ &  & 2.73  & 2.82  &  & 3.17  &  & \\
\hline\hline
\end{tabular}
\end{table}

\subsection{Excited states of $^8$He}
\begin{figure}[th]
\epsfxsize=8. cm
\centerline{\epsffile{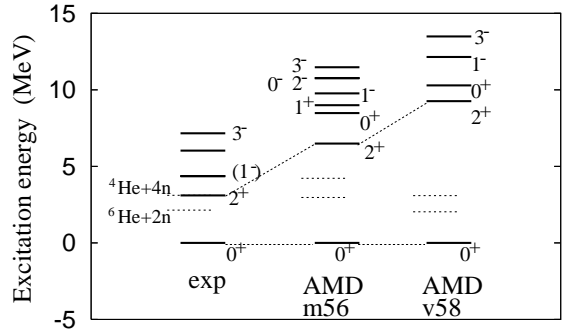}}
\vspace*{8pt}
\caption{Energy levels of $^8$He. The calculated results are those  
with the m56 and v58 interactions.
The experimental data are taken from \protect\cite{nucldata}.
\label{fig:he8spe}
}
\end{figure}

The calculated energy levels of $^8$He are illustlated 
in Fig.~\ref{fig:he8spe}, 
and the
properties of the excited states are shown in table \ref{tab:he8ex}.
In both of the m56 and v58 results,
the $2^+_1$ state is the lowest excited state and the $0^+_2$ state appears 
just above the $2^+_1$ state. The $1^-_1$ and $3^-_1$ states are obtained
in a higher energy region. 
In addition, in the present calculations with the m56 interaction, 
the $1^+_1$, $0^-_1$ and $2^-_1$ states are obtained
in almost the same energy region as the $1^-_1$ and $3^-_1$ states.
The present AMD framework is regarded as a kind of bound state approximation
because of the restricted model space, and therefore, 
coupling with continuum states is not taken into account.
In such a case, only resonance states remain in low-energy region
while continuum states rise to a high excitation energy region in principle. 
However, in order to check the stability of the resonances 
against neutron decays, their properties should be carefully examined.
In the present m56 results, the negative-parity states 
contain large component of 
$^6$He+$n+n$-like configurations with the valence neutron far from the
core. Since they have extremely large neutron radii and 
show somehow escaping behavior of neutrons,  
further investigation is required for these negative-parity states.
In particular, the $1^-_1$, $2^-_1$ and $0^-_1$ states can couple 
with $(0s)^2(0p)^3(1s)^1$ neutron configuration which has a 
valence $1s_{1/2}$ neutron with no centrifugal barrier.

\begin{table}[ht]
\caption{ 
\label{tab:he8ex} Excitation energies, Root-mean-square radii of point-proton, point-neutron and point-matter density, the expectation values of 
squared total intrinsic spin of neutrons $\langle S_n^2 \rangle$,
 and those of
the spin-orbit force $\langle V_{ls} \rangle$.
}
\begin{tabular}{cc|c|cccccc|cccccc}
\hline\hline
 & &   exp.  &AMD-v58 &&&&&&AMD-m56&&&&& \\          
nucleus & $J^\pi_n$ & $E_x$ & $E_x$ & $r_p$ & $r_n$ & $r_m$ & $\langle S^2_n \rangle$ & $\langle V_{ls} \rangle$ & $E_x$ & $r_p$ & $r_n$ & $r_m$ & $\langle S^2_n \rangle$ & $\langle V_{ls} \rangle$ \\
 &  & (MeV) & (MeV) & (fm) & (fm) & (fm) &  & (MeV) & (MeV) & (fm) & (fm) & (fm) &  & (MeV) \\
\hline
$^6$He & $2^+_1$ & 1.797 & 3.2  & 1.82 & 2.42 & 2.23& 0.19 & $-$2.3  & 2.6  & 1.87  & 2.46  & 2.28  & 0.27  & $-$2.3 \\
$^6$He & $0^+_1$ & 0 & 0.0  & 1.83  & 2.40  & 2.23  & 0.16  & $-$2.6  & 0.0  & 1.90  & 2.49  & 2.31  & 0.26  & $-$2.3 \\
\hline                            
$^8$He & $0^-_1$ &  &  &  &  &  &  &  & 10.8  & 2.13  & 3.63  & 3.32  & 2.05  & $-$5.9 \\
$^8$He & $2^-_1$ &  &  &  &  &  &  &  & 10.8  & 2.07  & 3.41  & 3.13  & 2.00  & $-$6.2 \\
$^8$He & $1^+_1$ &  &  &  &  &  &  &  & 9.0  & 1.94  & 2.81  & 2.62  & 2.03  & $-$2.5 \\
$^8$He & $3^-_1$ & 7.16 & 13.5  & 1.90  & 2.89  & 2.68  & 0.64  & $-$6.7  & 11.5  & 2.09  & 3.31  & 3.05  & 1.02  & $-$5.3 \\
$^8$He & $1^-_1$ & 4.36 & 12.1  & 1.95  & 3.05  & 2.82  & 0.81  & $-$7.9  & 9.8  & 2.13  & 3.52  & 3.23  & 1.24  & $-$5.8\\ 
$^8$He & $0^+_2$ &  & 10.3  & 1.97  & 2.94  & 2.73  & 0.67  & $-$4.7  & 8.5  & 2.11  & 3.12  & 2.90  & 0.99  & $-$1.0 \\
$^8$He & $2^+_1$ & 3.1 & 9.3  & 1.76  & 2.48  & 2.32  & 0.39  & $-$4.8  & 6.5  & 1.93  & 2.65  & 2.49  & 0.40  & $-$2.8 \\
$^8$He & $0^+_1$ & 0 & 0.0  & 1.76  & 2.37  & 2.24  & 0.72  & $-$11.4  & 0.0  & 1.96  & 2.63  & 2.48  & 0.86  & $-$7.3 \\
\hline                            
$^{10}$He & $0^+_1$ & 0 & 0.0  & 2.04  & 2.88  & 2.73  & 0.13  & $-$2.6  & 0.0  & 2.13  & 2.97  & 2.82  & 0.11  & $-$1.7 \\
\hline\hline
\end{tabular}
\end{table}

Compared with the experimental data, 
the theoretical values of the $2^+_1$ excitation energy 
are higher than the experimental one.
However, it is important that the level structure for the excited states,
$2^+_1$, $0^+_2$, $1^-_1$ and $3^-_1$, is not sensitive
to the adopted interaction  
though the relative position to the ground energy
depends on the interaction. 
The $0^+_2$ state is theoretically suggested to appear just above the $2^+_1$ 
state. What is striking is that 
the $0^+_2$ state has a remarkably large neutron radius compared
with the ground state because of developed $^4{\rm He}+2n+2n$ structure.
In the obtained wave function of the $0^+_2$ state, which is 
given by a superposition of the basis AMD wave functions,
the amplitude is found to be 
widely distributed into the basis wave functions with
various spatial configuration of $^4{\rm He}+2n+2n$. 
This indicates a gas-like
feature that the dineutrons are rather freely moving around the $^4$He core.
Therefore, 
we consider that
the $0^+_2$ state is the candidate of the cluster gas-like state with two
dineutrons around the $\alpha$ core.
The detailed discussion of the dineutron-like structure is given later. 
In the experimental energy spectra, 
some excited states were observed above the 
$2^+_1$ state. Spins and parities of these states are not 
definitely assigned yet. In the present calculations, the predicted 
$0^+_2$ state has the strong monopole neutron transition from the ground 
states as the matrix element 
$M_n(0^+_1\rightarrow 0^+_2)=13.5(13.9)$ fm$^2$ in the m56(v58) results.
This neutron matrix element is much larger than the 
observed proton matrix element $M_n(0^+_1\rightarrow 0^+_2)=5.4$ fm$^2$
of $^{12}$C by more than factor 2.
Therefore, we consider that the $^8$He($0^+_2$)  
might be excited by inelastic scattering on nuclear target.  

The excited states of $^8$He have been theoretically predicted 
by a few other calculations such as 
CSM and GFMC. 
The CSM gives better agreement of the 
$2^+_1$ excitation energy with the experimental 
data\cite{Volya05}.
We also comment that the GFMC calculation with AV18/IL2, which is 
an {\it ab initio} calculation with 
the realistic 2-body force and the empirical 3-body force, gives similar
level structure to the present m56 results.
Namely, the GFMC with AV18/IL2 gives
the $2^+$ state at $E_x=4.72$ MeV
and the $1^+_1$, $0^+_2$ and $2^+_2$ 
states in the $E_x >5 $ MeV region.

\section{Dineutron structure} \label{sec:discussions}

\subsection{What is dineutron($^2n$) cluster ?}
There is no bound state in an isolate $nn$ system. However, 
it has been emphasized in many theoretical works 
that
the spatial neutron-neutron correlation 
plays an important in the binding mechanism 
of the Borromean systems with two-neutron halo such as 
$^6$He and $^{11}$Li
(for example, \cite{Bertsch91,Zhukov93,Aoyama01,Aoyama94} 
and references therein).
The neutron-neutron correlation is characterized by 
a spin-zero $nn$ pair with spatial correlation in $S$ wave. 
In the correlation density of two-neutron halo nuclei, 
a peak of the probability appears at the region with 
a small $n$-$n$ distance($R(nn)$) and a large $n$-core distance 
in general. This corresponds to the dineutron correlation. 
In an extended meaning, it is regarded as 
a ``dineutron cluster'' which can virtually exist 
in loosely bound neutron-rich nuclei. 

As mentioned above, the characteristics of the dineutron 
are the zero spin and the spatial correlation. 
In the correlation density for $^6$He, $^{11}$Li and $^{14}$Be given 
by three-body calculations
\cite{Zhukov93,Arai01,Descouvemont-he6,Descouvemont-be14}, 
the peak for the dineutron correlation are 
seen typically around the $R(nn)\sim =2$ fm 
with a ridge in the $R(nn)=2\sim 3$ fm region.
It is important that this $n$-$n$ distance at the peak nearly depends on the 
system size among these three systems, $^6$He, $^{11}$Li and $^{14}$Be.
From this most probable $n$-$n$ distance, 
the typical size of the spatial correlation 
of the $nn$ pair can be estimated to be about 2 fm.
Then, we here approximately describe the dineutron cluster, $^2n$, 
by a spin-zero neutron pair written by the simple harmonic-oscillator 
$(0s)^2$ state with the size parameter $b$ in order to 
investigate dineutron structure in $^8$He. Then, the $^2n$-cluster
wave function $\phi^{^2n}({\bf S})$ which is localized at the position 
${\bf S}$ is expressed as,
\begin{eqnarray}
&\phi^{^2n}({\bf S})={\cal{A}}\left\{\phi^{0s}_{{\bf S}}
({\bf r}_1) \chi_\uparrow 
\phi^{0s}_{{\bf S}}({\bf r}_2)\chi_\downarrow \right\},\\
& \phi^{0s}_{\bf S}({\bf r}_i) = \frac{1}{(b^2\pi)^{\frac{3}{4}}} 
\exp\bigl\{-\frac{1}{2b^2}({\bf r}_i-{\bf S})^2\bigr\}.
\end{eqnarray} 
In this definition, 
the relative motion between two neutrons in the $^2n$ cluster is given 
by a Gaussian,
\begin{equation}
 \phi^r({\bf r}_1-{\bf r}_2) = \frac{1}{(b_r^2\pi)^{\frac{3}{4}}} 
\exp\bigl\{-\frac{1}{2b_r^2}({\bf r}_1-{\bf r}_2)^2\bigr\},
\end{equation}
with the size $b_r=\sqrt{2}b$, which should be 
the typical $nn$ distance $b_r=2\sim 3$ fm. With this approximation of the
$^2n$ cluster, major component of the dineutron correlation might be
taken into account, though the tail part at the large correlation length 
is omitted. For simplicity, we chose 
the size parameter $b$ for the $(0s)^2$ dineutron cluster as 
$b=1/\sqrt{2\nu}$, where $\nu$ is the width parameter $\nu(^6$He) 
and $\nu(^8$He) optimized 
for the $^6$He and $^8$He, respectively, in the AMD calculations.
The values $\nu$, which are listed in table \ref{tab:int},
 correspond to $b_r=2.0-2.3$ fm and   
satisfy the typical $nn$ distance of the dineutron correlation. 

\subsection{dineutron-cluster motion}

\begin{figure}[th]
\epsfxsize=7. cm
\centerline{\epsffile{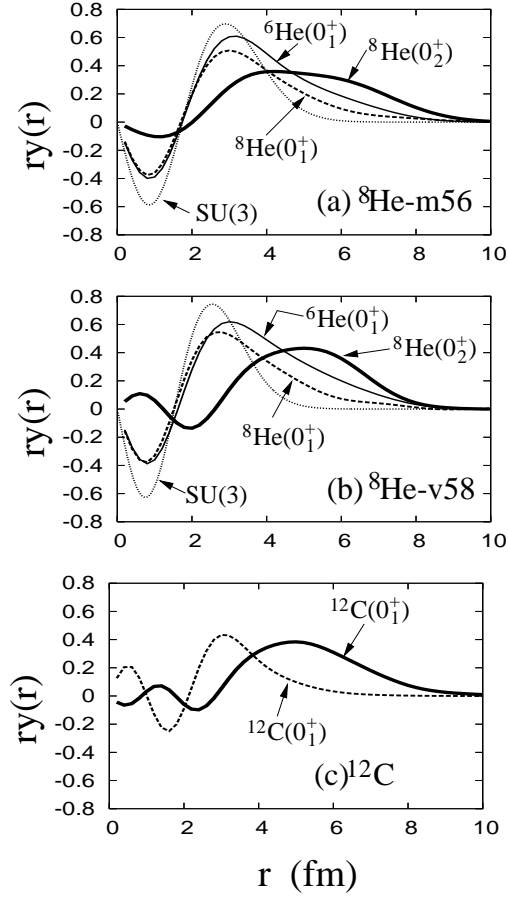}}
\vspace*{8pt}
\caption{Reduced width amplitudes $ry_{l=0}(r)$ for 
$^6$He$^{SU(3)}(0^+)$-$^2n$ in the $^8$He($0^+$), and those for 
$^4$He-$^2n$ in the $^6$He($0^+$). The $^8$He($0^+$) and $^6$He($0^+$)
wave functions are calculated by AMD+GCM with (a) the v58 and (b) the 
m56 interactions. The $^6$He$^{SU(3)}(0^+)$ is written by a
SU(3)-limit $^4$He-$2n$ cluster state. 
The $^4$He cluster and the $^2n$ cluster
are expressed by the $(0s)^4$ and $(0s)^2$ wave functions, respectively, 
where the size parameter for the $(0s)$ state  
is chosen to be the same value 
as the AMD+GCM wave functions; 
$b=1/\sqrt{2\nu(^6{\rm He})}$ in the calculation of $ry(r)$ 
for $^6$He($0^+$) and $b=1/\sqrt{2\nu(^8{\rm He})}$ in the calculation of 
$ry(r)$ for $^8$He($0^+$).
The reduced width amplitudes for $^4$He-$^2n$ in the $^6$He$^{SU(3)}(0^+)$
are also shown. (c) The reduced width amplitudes for 
$^8$Be$^{SU(3)}(0^+)$-$\alpha$ in the $^{12}$C$(0^+)$ taken from 
\protect\cite{Enyo-c12v2}.
\label{fig:yl}
}
\end{figure}

In order to investigate features of dineutron cluster 
structure in the $0^+$ states 
of $^8$He, we extracted the $^2n$-cluster motion from the obtained 
$^8$He($0^+$) wave functions. We assume a simple core
$(^4{\rm He}+^2n)_{0^+}$ which is equivalent to the SU(3)-limit $^6$He($0^+$),
and form the $^6$He$^{SU(3)}(0^+)$-$^2n$ cluster wave function 
with the $L=0$ relative motion between the 
core $^6$He$^{SU(3)}(0^+)$ and the $^2n$ cluster. 
In the same way as \cite{Enyo-be12,Enyo-c12v2} 
for $\alpha$-cluster motion,
we calculated the reduced width amplitudes $ry(r)$ 
for the $^2n$-cluster motion
and the cluster probability $S^{\rm fac}$ by taking the overlap of the 
$^6$He$^{SU(3)}(0^+)$-$^2n$ cluster wave functions with the $^8$He wave 
functions. 
In Fig.~\ref{fig:yl}, we show the reduced width amplitudes
in the $^8$He$(0^+_1)$ and the $^8$He$(0^+_2)$ wave functions
 obtained by the v58 and m56
interactions. These indicate the $^6$He$^{SU(3)}(0^+)$-$^2n$ relative motion.
 We also show the reduced width amplitudes for 
the $^8$Be$^{SU(3)}(0^+)-\alpha$ relative motion
in the $^{12}$C($0^+_1)$ and $^{12}$C($0^+_2)$ given in \cite{Enyo-c12}.
Surprisingly, the $^2n$-cluster motion in the $^8$He is quite similar to the 
$\alpha$-cluster motion in the $^{12}$C. 

First we discuss the features of the 
dineutron clustering in the $0^+_2$ state.
The most striking thing is that the $^8$He$(0^+_2)$ state has 
the large amplitude of the dineutron cluster 
in the long distance region around $r=4-6$ fm, which well corresponds
to the peak position of the $\alpha$-cluster motion in the
$^{12}$C($0^+_2$). 
The enhancement of the $^2n$-cluster component at the long distance
is more remarkable in the v58 results than the m56 results. 
The cluster probability of the $^8$He($0^+_2$), which is defined by the
integrated overlap with the $^6$He$^{SU(3)}(0^+)$-$^2n$ cluster wave functions,
is $S^{\rm fac}=0.50$ and $S^{\rm fac}=0.43$ in the v58 
and the m56 results. The larger development of the $^2n$ 
clustering in the v58 results is considered to be because of 
the stronger $n$-$n$ interaction in the v58 than the m56 interaction. 
It is very important that, 
even with the weaker $n$-$n$ interactions of the m56, 
the $^2n$-cluster structure 
survives with the significant component 
in the $^8$He($0^+_2$). 
Considering that the other $^2n$ cluster exists 
inside the $^6$He$^{SU(3)}(0^+)$ core, 
it is regarded that the $^8$He($0^+_2$) has the component of the
developed $^4$He+$^2n$+$^2n$ clustering, where two dineutrons are moving
in $L=0$ orbits. 
Furthermore, from the analogy of the $^2n$-cluster structure 
in the $^8$He($0^+_2$)
with the $\alpha$-cluster structure in the $^{12}$C,
the $^8$He($0^+_2$) is considered to contain the dineutron gas-like structure.

Next, we discuss dineutron structure in the ground state of $^8$He.
In the $^8$He($0^+_1$), 
the reduced width amplitude
has a peak at the distance less than 3 fm. It means that the
spatial development of the $^2n$ cluster is not so remarkable as
that of the $^8$He$(0^+_2)$. 
After discussing dineutron structure in
the $^6$He($0^+_1$), we shall 
compare it with the dineutron structure in the $^8$He($0^+_1$).
In Fig.~\ref{fig:yl}, we show the reduced width amplitudes 
of the $^4$He-$^2n$ cluster 
motion in the $^6$He($0^+_1$) obtained by the present calculations,
and that in the $^6$He$^{SU(3)}(0^+)$ given by 
the SU(3)-limit $^4$He-$^2n$ state.
Compared with the SU(3)-limit, the calculated 
$^6$He($0^+_1$) wave function has a long tail of dineutron structure 
at the surface. The $^2n$-cluster probability in the 
$^6$He($0^+_1$) state is $S^{\rm fac}=0.91$ and 0.84 in the v58 and the
m56 calculations. This is consistent with the fraction, 0.92 and 0.87,  
of the $S=0$ component, which are 
estimated from $\langle S^2_n \rangle $.
The $^2n$-cluster probability is reduced by the $S=1$ component 
because of the mixing of the $(p_{3/2})^2$ state. 
The dineutron wave function in the inner region 
is similar to that of 
the SU(3)-limit $^4$He-$^2n$ state. 
In this region, we have better to call it 
the spin-zero $2n$ correlation(dineutron correlation) rather than 
the $^2n$ cluster,
 because the antisymmetrization effect is important there. 

Comparing the result of $^8$He$(0^+_1)$ with 
that of $^6$He($0^+_1$), we found that the reduced width amplitude for the 
dineutron component is suppressed 
in the $^8$He$(0^+_1)$. This is because of the $p_{3/2}$ sub-shell closure
effect. As mentioned in the previous section, 
the $j$-$j$ coupling feature is more remarkable 
in the $^8$He$(0^+_1)$ than the
$^6$He($0^+_1$).
However, the cluster probability of the $^8$He$(0^+_1)$ is still significant as
$S^{\rm fac}=0.57$ and 0.52 in the v58 and the m56 results, respectively.
This probability dominantly originates in 
the SU(3)-limit $^4$He+$^2n$+$^2n$ configuration,
 which is equivalent to the $L$-$S$ coupling 
$p$-shell configuration. It means that the dineutron correlation
is still important in the $^8$He($0^+_1$).
This situation is quite similar to that of the $^{12}$C$(0^+_1)$ which is 
the admixture of the $p_{3/2}$ closure and 
the SU(3)-limit $3\alpha$ state.
As a result of the $L$-$S$ coupling feature due to the 
dineutron correlation, the  
$^8$He$(0^+_1)$ state should contain the significant 
$(p_{3/2})^2(p_{1/2})^2$ contamination. This result is consistent with the
experimental indication of the  
$p_{1/2}$ component in the $^8$He ground state
reported by the recent observations\cite{Chulkov05,Keeley07}.
As seen in Fig.~\ref{fig:yl}, it is also interesting that
the $^8$He($0^+_1$) state has a tail of the 
$^2n$-cluster motion at the surface, though the tail is slight compared
with the long tail in the $^6$He($0^+_1$).
In conclusion, the $^8$He($0^+_1$) is the admixture of
the $p_{3/2}$ closure and the $L$-$S$ coupling $p$-shell configuration 
of neutrons with a small tail of the dineutron clustering.

\subsection{$^2n$ condensate wave function}
In the previous subsection, we discuss the $^2n$-cluster wave function by 
assuming the core $(^4{\rm He}+^2n)_{0^+}$ 
which is equivalent to the SU(3)-limit $^6$He($0^+$).
In this description, one of the $^2n$ clusters is confined in the 
the core $(^4{\rm He}+^2n)_{0^+}$, and its relative 
wave function to the $^4$He is given by the $1s$ orbit 
of the harmonic oscillator potential with 
the oscillator frequency $\omega=8\nu/3$. 

As shown in Fig.~\ref{fig:yl}, in this SU(3)-limit, 
the radial wave function of the $^2n$-cluster around the $^4$He 
remains in the inner region. In such the case, 
although the $^2n$-cluster is moving in the $S$ wave,
the $^2n$-cluster receives
much effect of antisymmetrization from the $^4$He core 
and it does not necessarily indicate a gas-like state.
In order to see more directly the $^2n$-cluster gas-like nature,
where two $^2n$'s are moving in $S$ wave far from the 
the $^4$He core,  
we assumed the $^2n$ condensate wave function in the 
$^4{\rm He}+^2n+^2n$ system and calculated the overlap with 
the obtained $^8$He($0^+$) wave functions.

We define the $^2n$ condensate wave function by naturally extending
the $\alpha$ condensate wave function proposed by 
Tohsaki et al.\cite{Tohsaki01}
as follows,
\begin{equation}
\Psi_{\rm cond}(B)\equiv n_0 \int \prod^k_{i=1} \left\{ d^3{\bf S}_i 
\exp\left( -\frac{({\bf S}_i-{\bf S}_C)^2}{B^2}\right ) \right\}
\Phi_{\rm Brink}({\bf S}_C,{\bf S}_1,{\bf S}_2,\cdots
{\bf S}_k),
\end{equation}
where $n_0$ is the normalization factor and $\Phi_{\rm Brink}({\bf S}_C,{\bf S}_1,{\bf S}_2,\cdots
{\bf S}_k)$ is the Brink wave function for the $C+k(^2n)$-cluster
 system consisting the core($C$) and 
$k$ dineutrons($^2n$) as,
\begin{equation}
\Phi_{\rm Brink}({\bf S}_C,{\bf S}_1,{\bf S}_2,\cdots
{\bf S}_k)\equiv {\cal{A}}\left\{\phi^{\rm C}({\bf S}_C)\phi^{^2n}({\bf S}_1)
\phi^{^2n}({\bf S}_2)\cdots \phi^{^2n}({\bf S}_k) 
\right \}.
\end{equation}
Here, the wave function of the $i$th $^2n$, 
$\phi^{^2n}({\bf S}_i)$, is given by the $(0s)^2$ state localized around
${\bf S}_i$. ${\bf S}_C$ is the mean position of the 
center of mass motion of the core, and is chosen to be
${\bf S}_C=-\frac{2}{A}({\bf S}_1+{\bf S}_2+\cdots+{\bf S}_k)$.
In heavy limit of the core mass $A$, this wave function is equivalent
to the dineutron condensate wave function 
proposed by Horiuchi\cite{Horiuchi06}.
In the present calculation for 
 $^4{\rm He}+^2n+^2n$, the core $C$ is $^4$He, and 
the number of $^2n$ clusters is $k=2$. We assumed the $(0s)^4$ state
of the core wave function, $\phi^{^4{\rm He}}$, and adopted the common 
size parameter $b=1/\sqrt(2 \nu(^8{\rm He})$ for the 
$^4{\rm He}$ and $^2n$ clusters.
In the practical calculations, the 6-dimensional integrals for the coordinates,
${\bf S}_1$ and ${\bf S}_2$, are 
performed by taking mesh points on $(\theta_{12},|{\bf S}_1|,|{\bf S}_2|)$
and the total-angular-momentum projection
($\theta_{12}$ is the angle between ${\bf S}_1$ and ${\bf S}_2$).

\begin{figure}[th]
\epsfxsize=6. cm
\centerline{\epsffile{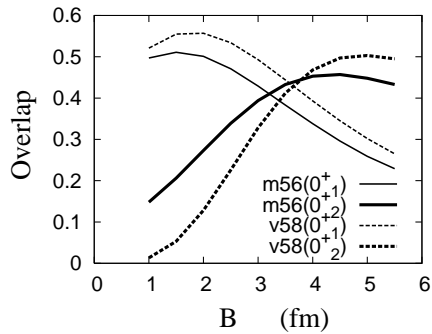}}
\vspace*{8pt}
\caption{The squared overlap between the dineutron 
condensate wave function $\Psi_{\rm cond}(B)$ and the obtained
$^8$He$(0^+)$ wave functions. See details in the text.
\label{fig:bec}
}
\end{figure}

In Fig.~\ref{fig:bec}, we show the squared overlap, 
$|\langle ^8{\rm He}|\Psi_{\rm cond}(B)\rangle|^2$, between 
the $^2n$ condensate wave function and 
the $^8$He wave functions obtained by AMD+GCM. 
The calculated values are plotted as a function $B$ which indicates the
size of the spatial distribution of $^2n$ clusters in the 
condensate wave function.
The $^8$He$(0^+_1)$ has the overlap, about 0.5, 
at $B < 2$ fm. The condensate wave function 
$\Psi_{\rm cond}(B)$ with such a small size $B$ is almost equivalent to
the SU(3)-limit $^4$He+$2n$+$2n$ state. 
On the other hand, 
the $^8$He$(0^+_2)$ has the maximum overlap, about 0.5, at 
remarkably large size $B=4-5$ fm. 
This is an strong indication of the dineutron gas-like component
in the calculated $^8$He$(0^+_2)$. 
The dineutron gas-like feature is further enhanced 
in case of the v58 interaction 
than the m56 interaction. 
These results are consistent with the discussion 
of the $^2n$-cluster wave function 
in the previous subsection.  

\section{Summary}\label{sec:summary}

We studied the structure of $^8$He with a method of AMD+GCM.
We chose the effective nuclear interactions
by taking care of energies of subsystems, and reproduced the 
properties of ground states of $^4$He, $^6$He and $^8$He.
In the ground state of $^8$He, the component of the 
$p_{3/2}$ sub-shell closure is dominant. However, 
the $L$-$S$ coupling feature is also significantly contained  
because of the spin-zero dineutron correlation. This is consistent with the
experimental report on the significant mixing of 
$(p_{3/2})^2(p_{1/2})^2$ component in the $^8$He($0^+_1)$.
It is concluded that the $^8$He($0^+_1)$ is 
the admixture of $p_{3/2}$ sub-shell closure
and $L$-$S$ coupling $p$-shell configurations with a slight dineutron tail
at the surface. This result is also consistent with the experimentally
suggested large spectroscopic factor of the $^6$He($2^+$) 
in the $^8$He($0^+_1)$.

The present results suggest that the $0^+_2$ state may appear a few MeV
above the $2^+_1$ state. By analyzing dineutron structure, 
it was found that this state has a significant component of 
the developed $^4$He+$^2n$+$^2n$ structure where two dineutrons are
moving around the $^4$He core in $S$ wave with a dilute density.
The $^2n$-cluster wave function of the $^8$He($0^+_2$) state 
is similar to the
$\alpha$-cluster wave function of the $^{12}$C($0^+_2$) state.
Therefore, we consider that the predicted $0^+_2$ state is the candidate of 
the dineutron gas-like state, which is analogy to the 
$\alpha$ condensate state suggested in the $^{12}$C($0^+_2$).
In the experimental energy spectra of $^8$He, 
some excited states were observed above the 
$2^+_1$ state. Spins and parities of these states have not been  
definitely assigned yet.
Since the present calculations
predicted the remarkable neutron matrix element for the monopole 
transitions $^8$He($0^+_1)\rightarrow ^8$He($0^+_2)$, 
we expect that the $^8$He($0^+_2$)  might be excited 
in inelastic scattering on nuclear target.  

Since the AMD framework is regarded as a kind of bound state approximation 
because of the restricted model space,
coupling with continuum states is not taken into account.
In future study, widths of the excited states should be carefully 
investigated 
by taking into account the continuum coupling in order 
to confirm the stability of the resonances against 
particle decays.

In the present work, the calculations were performed 
within the AMD model space 
by using effective interactions. 
We chose the interaction parameters
by taking care of subsystem energies such as 
$\alpha$-$n$, $^6$He as well as 
nucleon-nucleon systems. Although it is difficult to 
completely reproduce all of the subsystem energies with a unique 
effective interaction, we found the interaction which can 
reasonably reproduce the global feature of the subsystem energies.
We here stress that the level structure of the excited states 
is not sensitive to the 
adopted nuclear forces within the reasonable choice of effective
interaction, though the excitation energy relative to the ground state
depends on the interaction. It is also important that 
the dineutron structure of the $^8$He$(0^+$) states is qualitatively 
similar among four sets of 
interaction adopted in the present calculations.
For further investigations of He isotopes, more extended 
calculations based on the realistic forces should be important as well as 
{\it ab initio} calculations.

\section*{Acknowledgments}

The authors would like to thank Prof. Horiuchi, Prof. Tohsaki and their collaborators for valuable discussions. 
They are also thankful to members of 
Yukawa Institute for Theoretical Physics(YITP)
and Department of Physics in Kyoto University, especially Dr. Takashina for
fruitful discussions.
The computational calculations in this work were performed by the 
Supercomputer Projects 
of High Energy Accelerator Research Organization(KEK)
and also the super computers of YITP.
This work was supported by 
Grant-in-Aid for Scientific Research 
Japan Society for the Promotion of 
Science and a Grant-in-Aid for Scientific Research from JSPS.
It is also supported by the Grant-in-Aid for 
the 21st Century COE "Center for Diversity and Universality in Physics"
from MEXT.
Discussions in the RCNP workshops on cluster physics held in 2007
and those in the workshops YITP-W-06-17 
and YITP-W-07-01 held in YITP were helpful 
to initiate and complete this work.


\begin{thebibliography}{0}
\bibitem{Tohsaki01}
A. Tohsaki, H. Horiuchi, P. Schuck, and G. R\"opke, 
Phys. Rev. Lett. {\bf 87}, 192501 (2001).
\bibitem{Ropke98}
G. R\"opke, A. Schnell, P. Schuck, and P. Nozieres, Phys. Rev. Lett. {\bf 80},
3177 (1998).
\bibitem{Matsuo06}
M.Matsuo, Phys. Rev. C {\bf 73}, 044309 (2006).
\bibitem{Bertsch91}
G. F. Bertsch and H. Esbensen, Ann. Phys. (NY) {\bf 209}, 327
(1991).
\bibitem{Zhukov93}
M. V. Zhukov {\em et al.}, Phys. Rep. {\bf 231}, 151 (1993).
\bibitem{Aoyama01}
S. Aoyama, K. Kato, and  K. Ikeda, 
Prog. Theor. Phys. Suppl. {\bf 142}, 35 (2001).
\bibitem{Arai01}
K. Arai, Y. Ogawa, Y. Suzuki, and K. Varga, Prog. Theor. Phys. Suppl. 
{\bf 142}, 97 (2001).
\bibitem{Pieper04}
S. C. Pieper, R. B. Wiringa, and J. Carlson,
Phys. Rev. C {\bf 70}, 054325 (2004). 
\bibitem{Pieper05}
S. C. Pieper, Nucl. Phys. A {\bf 751}, 516c (2005).
\bibitem{Caurier06}
E. Caurier and P. Navr\'atil, Phys. Rev. C {\bf 73}, 021302(R) (2006).
\bibitem{Michel03}
N. Michel, W. Nazarewicz, M. Ploszajczak, and J. Okolowicz
Phys. Rev. C {\bf 67}, 054311 (2003).
\bibitem{Volya05}
A. Volya and V. Zelevinsky, Phys. ReV. Lett. {\bf 94}, 052501 (2005);
A. Volya and V. Zelevinsky, Phys. Rev. C {\bf 74}, 064314 (2006).
\bibitem{Hagen05}
G. Hagen, M. Hjorth-Jensen, and J. S. Vaagen, 
Phys. Rev. C {\bf 71}, 044314 (2005).
\bibitem{Sugahara96}
Y. Sugahara {\em et al.}, Prog. Theor. Phys. {\bf 96} 1165 (1996).
\bibitem{Csoto93}
A. Cs\'ot\'o, Phys. Rev. C {\bf 48}, 165 (1993).
\bibitem{Baye94}
D. Baye, M. Kruglanski and M. Vincke, Nucl. Phys. A {\bf 573}, 431 (1994).
\bibitem{Aoyama02}
S. Aoyama, Phys. Rev. Lett. {\bf 89}, 052501 (2002).
\bibitem{Suzuki90}
Y. Suzuki, and W. J. Ju, Phys. Rev. C {\bf 41} 736 (1990).
\bibitem{Varga94}
K. Varga, Y. Suzuki, and Y. Ohbayasi, Phys. Rev. C {\bf 50},  189 (1994)
\bibitem{Itagaki00}
N. Itagaki and S. Aoyama, Phys. Rev. C {\bf 61}, 024303 (2000).
\bibitem{Masui07}
H. Masui, K. Kat\=o, and K. Ikeda, Phys. Rev. C {\bf 75}, 034316(2007).
\bibitem{Dote00}
A. Dot\'e and H. Horiuchi, Prog. Theor. Phys. {\bf 103}, 261 (2000).
\bibitem{Aoyama06}
S. Aoyama, N. Itagaki, and M. Oi, Phys. Rev. C {\bf 74}, 017307 (2006).
\bibitem{Neff05}
T. Neff, H. Feldmeier, and R. Roth, Nucl. Phys. A {\bf 752}, 321c (2005).
\bibitem{Tanihata92}
I.Tanihata {\em et al.}, 
Phys. Lett. {\bf 289B}, 261 (1992).
\bibitem{Korsheninnikov03}
A. A. Korsheninnikov {\em et al.}
Phys. Rev. Lett. {\bf 90}, 082501 (2003).
\bibitem{Chulkov05}
L. V. Chulkov {\em et al.}, Nucl. Phys. A {\bf 759} 43 (2005).
\bibitem{Keeley07}
N. Keeley {\em et al.}, Phys. Lett. {\bf B646}, 222 (2007).
\bibitem{Skaza06}
F. Skaza {\em et al.}, Phys. Rev. C {\bf 73}, 044301 (2006).
\bibitem{Korsheninnikov93}
A. A. Korsheninnikov {\em et al.}, 
Phys. Lett. {\bf B 316}, 38 (1993).
\bibitem{ENYObc}
 Y. Kanada-En'yo, H. Horiuchi and A. Ono,
Phys. Rev. C {\bf 52}, 628 (1995);
 Y. Kanada-En'yo and H. Horiuchi,
Phys. Rev. C {\bf 52}, 647 (1995).
\bibitem{ENYOsup}
Y. Kanada-En'yo and  H. Horiuchi, Prog. Theor. Phys. Suppl.{\bf 142},
 205 (2001).
\bibitem{AMDrev}
Y. Kanada-En'yo, M. Kimura and H. Horiuchi, Comptes rendus Physique Vol.4, 
497 (2003).
\bibitem{Enyo-c12v2}
Y. Kanada-En'yo, Prog. Theor. Phys. {\bf 117}, 655 (2007).
\bibitem{Enyo-c11}
Y. Kanada-En'yo, 
Phys. Rev. C {\bf 75}, 024302 (2007).
\bibitem{Kimura04}
M.Kimura and H.Horiuchi,
Prog. Theor. Phys. {\bf 111}, 841 (2004).
\bibitem{Enyo04}
Y.Kanada-En'yo and Y.Akaishi,
Phys.Rev. C 69, 034306 (2004)
\bibitem{Dote97}
A.Dote, H.Horiuchi, and Y.Kanada-En'yo,
Phys.Rev. C56, 1844 (1997).
\bibitem{Volkov}
 A. B. Volkov, Nucl. Phys {\bf 74}, 33 (1965).
\bibitem{MV1}
 T. Ando, K. Ikeda and A. Tohsaki,
        Prog. Theory. Phys. {\bf 64}, 1608  (1980).
\bibitem{LS}
 N. Yamaguchi, T. Kasahara, S. Nagata and Y. Akaishi,
 Prog. Theor. Phys. {\bf 62}, 1018 (1979);
 R. Tamagaki, Prog. Theor. Phys. {\bf 39}, 91 (1968).
\bibitem{Enyo-c12}
 Y. Kanada-En'yo,
Phys. Rev. Lett. {\bf 81}, 5291 (1998).
\bibitem{Tanihata85}
I. Tanihata {\em et al.},
Phys. Rev. Lett. {\bf 55}, 2676 (1985).
\bibitem{Tanihata88}
I. Tanihata {\em et al.},
Phys. Lett. B {\bf 206}, 592 (1988).
\bibitem{Hagino05}
K. Hagino and H. Sagawa,
Phys. Rev. C {\bf 72}, 044321 (2005).
\bibitem{Wang04}
L. -B. Wang {\em et al.}
Phys. Rev. Lett. {\bf 93}, 142501 (2004).
\bibitem{nucldata}
D. R. Tilley {\em et al.}, Nucl. Phys. A {\bf 745}, 155 (2004).
\bibitem{Aoyama94}
S. Aoyama, A. Muraki, K. Kat\=o, and K. Ikeda,
Prog. Theor. Phys. {\bf 93}, 99 (1995).
\bibitem{Descouvemont-he6}
P. Descouvemont, C. Daniel, and D. Baye,
Phys. Rev. C {\bf 67}, 044309 (2003).
\bibitem{Descouvemont-be14}
P. Descouvemont, E. Tursunov, and D. Baye,
Nucl. Phys. {\bf A765}, 370 (2006).
\bibitem{Enyo-be12}
Y. Kanada-En'yo and H. Horiuchi, Phys. Rev. C {\bf 68}, 014319 (2003).
\bibitem{Horiuchi06}
H. Horiuchi, Mod. Phys. Lett. {\bf A21}, Nos.31-33, 2455 (2006).







\end{thebibliography}
\end{document}